\documentclass[12pt,draftcls,journal,onecolumn]{IEEEtran}
\usepackage{graphicx}
\usepackage{epstopdf}
\usepackage[latin1]{inputenc}
\usepackage{amsmath,amssymb,amscd,latexsym,dsfont}
\usepackage[english]{babel}
\usepackage{tabularx,cite}
\usepackage{graphicx}
\usepackage{algpseudocode}
\usepackage{algorithm}
\usepackage{algorithmicx}
\usepackage{float}
\usepackage{multicol}
\usepackage{psfrag}
\usepackage{comment,psfrag,subfigure,enumerate}
\usepackage{color}
\usepackage{amsthm}
\usepackage{graphicx}
\usepackage{graphicx}
\usepackage{epstopdf}
\usepackage[latin1]{inputenc}
\usepackage{amsmath,amssymb,amscd,latexsym,dsfont}
\usepackage[english]{babel}
\usepackage{tabularx,cite}
\usepackage{graphicx}
\usepackage{algpseudocode}
\usepackage{algorithm}
\usepackage{algorithmicx}
\usepackage{float}
\usepackage{multicol}
\usepackage{mathtools}
\usepackage{psfrag}
\usepackage{comment,psfrag,subfigure,enumerate}
\usepackage{color}
\usepackage{amsthm}
\ifCLASSINFOpdf
\else
\fi
\begin{document}
\IEEEoverridecommandlockouts
\IEEEpubid{\makebox[\columnwidth]{978-1-4799-5863-4/14/\$31.00 \copyright 2014 IEEE \hfill} \hspace{\columnsep}\makebox[\columnwidth]{ }}
\title{Performance Analysis of ARQ-based RF-FSO Links}

\author{\IEEEauthorblockN{Behrooz Makki, Tommy Svensson, Thomas Eriksson and Mohamed-Slim Alouini, \emph{Fellow, IEEE}}\\
%\IEEEauthorblockA{Department of Signals and Systems,
%Chalmers University of Technology, Gothenburg, Sweden\\
%Email: \{behrooz.makki, alexandre.graell, thomase\}@chalmers.se}
%%\and
%%\IEEEauthorblockN{Thomas Eriksson}
%%\IEEEauthorblockA{Department of signals and systems\\
%%Chalmers University of Technology\\
%%Gothenburg, Sweden\\
%%Email: thomase@chalmers.se}
%\thanks{Alexandre Graell i Amat was supported by the Swedish Agency for Innovation Systems (VINNOVA) under the P36604-1 MAGIC project.}
\thanks{Behrooz Makki, Tommy Svensson and Thomas Eriksson are with Chalmers University of Technology, Email: \{behrooz.makki, tommy.svensson, thomase\}@chalmers.se. Mohamed-Slim Alouini is with the King Abdullah University of Science and Technology (KAUST), Email: slim.alouini@kaust.edu.sa}
}

%Telephone: (800) 555--1212, Fax: (888) 555--1212}


\maketitle
%\onecolumn
\vspace{-8mm}
\begin{abstract}
In this letter, we study the performance of hybrid radio-frequency (RF) and free-space optical (FSO) links using automatic repeat request (ARQ). We derive closed-form expressions for the message decoding probabilities, throughput, and outage probability with different relative coherence times of the RF and FSO links. We also evaluate the effect of adaptive power allocation between the ARQ retransmissions on the system performance. The results show that joint implementation of the RF and FSO links leads to substantial performance improvement, compared to the cases with only the RF or the FSO link.
\end{abstract}
%Finally, we derive closed-form expressions for the maximum achievable throughput of the return-link satellite systems using optimal schedulers.

%Then, compared to open-loop communication setups, the implementation of power-adaptive ARQ reduces the average power by ? and ? dB, if a maximum of 2 and 3 retransmissions is utilized, respectively.


% \fi
\IEEEpeerreviewmaketitle
\vspace{-8mm}
\section{Introduction}
%The next generation of communication networks must provide high-rate reliable data streams.
Free-space optical (FSO) systems provide fiber-like data rates through the atmosphere using lasers or light emitting diodes and, consequently, are very promising to provide high-rate communication for the next generation of wireless networks \cite{6168189,6857378,5380093,5557647}.
%Thus, the FSO can be used for a wide range of applications such as last-mile access, fiber back-up, back-haul for wireless cellular networks, and disaster recovery.
However, such links are highly susceptible to atmospheric effects and, therefore, are unreliable. For this reason, the FSO link is sometimes combined with an additional radio-frequency (RF) link to create a hybrid RF-FSO setup.

To achieve data rates comparable to those in the FSO link, a millimeter wavelength carrier is typically selected for the RF link. As a result, the RF link is also subject to atmospheric effects such as rain. However, the good point is that these links are complementary because the RF (resp. the FSO) signal is severely attenuated by the rain (resp. the fog/cloud) while the FSO (resp. the RF) signal is not. Therefore, the link reliability and the service availability are considerably improved via joint RF-FSO based data transmission. On the other hand, automatic repeat request (ARQ) is a well-established approach to increase the link reliability. In this perspective, it is interesting to analyze the performance of RF-FSO systems using ARQ.

There are different works on the performance of RF-FSO links. In, e.g., \cite{Hamzeh,6364576}, the RF and the FSO links work separately and the RF link acts as a backup for the FSO link. On the other hand,  \cite{6503564,5342330,5351671,5427418} consider the case where the links work simultaneously. Finally, ARQ in RF (resp. FSO) systems is studied in, e.g., \cite{throughputdef,MIMOARQkhodemun,6918481} (resp. \cite{6168189,6857378,5380093,5557647}), while the ARQ-based RF-FSO systems have been rarely studied \cite{5427418,globecomalouini}.

This letter studies the RF-FSO links using ARQ. With different relative coherence times of the RF and FSO links, we derive closed-form expressions for the message decoding probabilities, throughput and outage probability. Also, we analyze the effect of adaptive temporal power allocation between the ARQ retransmissions on the system performance.

As opposed to \cite{Hamzeh,6364576}, we consider joint data transmission/reception in the RF and FSO links. Also, this letter is different from \cite{6503564,5342330,5351671,5427418,throughputdef,MIMOARQkhodemun,6918481,6168189,6857378,5380093,5557647} because we study the performance of ARQ in joint RF-FSO links and derive new analytical/numerical results on the message decoding probabilities, power allocation,  outage probability, and throughput which, to our best knowledge, have not been presented before. Finally, compared to our work \cite{globecomalouini}, this letter considers different ARQ protocol and channel models, and evaluates the effect of temporal power allocation on the system performance.

Our results show that depending on the relative coherence times of the links there are different suitable methods for the analysis of RF-FSO systems.
%2) Adaptive power allocation reduces the outage probability of the RF-FSO links considerably. Finally, 3)
Also, the joint implementation of the RF and FSO links leads to substantial performance improvement, compared to the cases with only the RF or the FSO link, particularly if adaptive power allocation is used.
%For instance, consider the exponential distribution and the common relative coherence times of the RF and FSO links. Then, with the initial code rate $R=5$ nats-per-channel-use (npcu), a maximum of $M=2$ retransmission rounds of the HARQ and the outage probability $10^{-2}$, the joint RF-FSO based data transmission reduces the required power by $16$ and $4$ dB, compared to the cases with only the RF or the FSO link, respectively (see Fig. 10 for more details).
\vspace{-4mm}
\section{System Model}
%\begin{figure}
%\centering
%  % Requires \usepackage{graphicx}
%  \includegraphics[width=0.7\columnwidth]{ZzZRFFSOideaFig1.eps}\\\vspace{-3mm}
%\caption{Channel model. The data is jointly transmitted by the RF and the FSO links and, in each round of HARQ, the receiver decodes the data based on all received signals.}\label{figure111}
%\end{figure}
We consider a joint RF-FSO system where the data sequence is encoded
into parallel FSO and RF bit streams. The FSO link employs intensity modulation and direct detection while the RF link modulates the encoded bits and up-converts the baseband signal to a millimeter wavelength RF carrier frequency. Then, the FSO and the RF signals are simultaneously sent to the receiver. At the receiver, the received RF (resp. FSO) signal is down-converted to baseband (resp. collected by an aperture and converted to an electrical signal via photo-detection) and the signals are sent to the decoder which decodes the received signals jointly.
%Finally, we assume perfect synchronization between the links.

The channel coefficients  are assumed to be known by the receiver, in harmony with \cite{throughputdef,MIMOARQkhodemun,6168189,6857378,6918481,5342330}. However, there is no feedback to the transmitter, except for the ARQ feedback bits. %The feedback channel is supposed to be delay- and error-free.
%Let us define a packet as the transmission of a codeword along with all its possible retransmissions.
As the most practical ARQ approach \cite{throughputdef}, we consider basic ARQ
%, also referred to as Type-I ARQ,
with a maximum of $M$ retransmissions.
%,i.e., the message is retransmitted a maximum of $M$ times.
Using basic ARQ, the scaled versions of the same codeword are sent in the successive retransmissions and the receiver disregards the previous messages, if received in error.
%Thus, the equivalent data rate, i.e., the code rate, at the end of round $m$ is $\frac{K}{mL}=\frac{R}{m}$ where $R=\frac{K}{L}$ denotes the initial code rate. In each round, the receiver combines all received sub-codewords to decode the message.
%We consider Type-I ARQ with a maximum of $M$
%retransmissions, i.e., the data is transmitted a maximum of $M$ times, and in each round the receiver disregards the previous messages, if received in error.
The retransmission continues until the message is correctly decoded or the maximum permitted retransmission round is reached. Note that setting $M=1$ represents the cases without ARQ.
\vspace{-3mm}
\section{Analytical results}
As shown in \cite{throughputdef,MIMOARQkhodemun}, for different channel models the throughput of ARQ protocols is given by
\vspace{-1mm}
\begin{align}\label{eq:eqeta1}
\eta=\frac{R\left(1-\Phi_M\right)}{\sum_{m=1}^{M}{\Phi_{m-1}}},
\end{align}
%\begin{align}\label{eq:eqoutprob1}
%\Pr(\text{Outage})=\Pr\left(W_M\le \frac{R}{M}\right),
%\end{align}
%
%\begin{align}\label{eq:eqfunc}
%\eta=\text{Function}(\Pr(W_m\le \frac{R}{m}),\forall m=1,\ldots,M)
%\end{align}
where $R$ (in nats per channel use (npcu)) is the code rate, $\Phi_m$ represents the probability that the data is not decoded correctly by the receiver in rounds $n=1,\ldots,m$ and $\Phi_0\doteq1.$ Also, the outage probability is given by $\Pr(\text{Outage})=\Phi_M.$
%respectively, where $W_m$ is the accumulated mutual information (AMI) at the end of round $m$ and $R$ denotes the initial code rate. Also, $\Pr(W_m\le \frac{R}{m})$ denotes the probability that the data is not correctly decoded up to the end of the $m$-th round.
%In this way, the throughput and the outage probability of ARQ protocols are monotonic functions of the probabilities $\Pr(W_m\le \frac{R}{m}),\forall m$. This is because the system performance depends on the retransmission round in which the codewords are correctly decoded. Moreover, the probability $\Pr(W_m\le \frac{R}{m})$ is directly linked to the AMI $W_{m}$ which is a random variable and function of the \textcolor{blue}{channel} realizations experienced in rounds $n=1,\ldots,m.$
Thus, to analyze the throughput and outage probability, the key point is to determine the
%AMIs as functions of channel realization(s) and find their corresponding cumulative distribution functions (CDFs)\footnote{The CDF and the probability distribution function (PDF) of a random variable $X$ are denoted by $F_X(.)$ and $f_X(.),$ respectively.} $F_{W_{m}},m=1,\ldots,M$
probabilities $\Phi_m,m=1,\ldots,M$. Then, having the probabilities, the considered performance metrics are obtained. For basic ARQ, in particular, we have
\vspace{-5mm}
\begin{align}\label{eq:probphim}
\Phi_m=\begin{cases}
\prod_{j=1}^m{\phi_j}  & \text{ if } m\ne 0, \\
 1 & \text{ if } m=0,
\end{cases}
\end{align}
where $\phi_j$ is the probability that the data is not decoded in round $j$. Here, (\ref{eq:probphim}) is based on the fact that 1) independent channel realizations are experienced in each round, and 2)
%a scaled version of the initial codeword is sent in each transmission of a packet and 3)
in each round, the receiver decodes the data only based on the received signal in that round.

To find $\phi_j$, we note that, as demonstrated by, e.g., \cite{5342330,598416}, in RF-FSO systems the RF link experiences very slow variations and the coherence time of the RF link is in the order of $10^{2}-10^{3}$ times larger than the coherence time of the FSO link. Here, we consider the setup
%as illustrated in Fig. 2
where the RF link remains constant during each round of ARQ \cite{throughputdef,MIMOARQkhodemun} while
%in each round of ARQ
$N$ different channel realizations are experienced in the FSO link. Then, we study two district cases with small (resp. large) values of $N$ which correspond to the scenarios with comparable (resp. considerably different) coherence times of the RF and FSO links. %Note that, while we analyze the cases with shorter coherence time of the FSO link, the same analysis holds for
Also, it is straightforward to extend the results to the cases with shorter coherence time of the RF link, compared to the coherence time of the FSO link.
 %and 2) as seen in Section IV.B, we can derive the results in the cases with few, possibly 1, \textcolor{blue}{channel} realizations of the FSO link during the packet transmission.

With the considered setup, we can use the results of \cite[Chapter 15]{4444444444} to find the probability $\phi_j$ as
\begin{align}\label{eq:eqWm1}\vspace{-7mm}
&\phi_j=\Pr\left(\log(1+P_{\text{RF},j}G_{\text{RF},j})+\mathcal{Y}_{(j,N)}<R\right),
\mathcal{Y}_{(j,N)}\doteq\frac{1}{N}{\sum_{i=1}^N{\log(1+P_{\text{FSO},j}G_{\text{FSO}, 1+(j-1)i})}}.
\end{align}
Here, $P_{\text{RF},j}$ and $P_{\text{FSO},j}$ are, respectively, the transmission powers of the RF and FSO links in the $j$-th round. Also, $G_{\text{RF},j}$ and $G_{\text{FSO},j}$'s denote the channel gains of the RF and FSO links, respectively.
\vspace{-3mm}
\subsection{Performance Analysis in the Cases with Considerably Different Coherence Times for the RF and FSO Links}
%Here, we consider the cases where the coherence times of the RF and FSO links are considerably different. Motivated by, e.g., \cite{5342330,1142964,598416,598420}, we concentrate on the cases with shorter coherence time of the FSO link, compared to the RF link. Meanwhile, the same analysis is valid, if the RF link experiences shorter \textcolor{blue}{coherence time} than the FSO link.

With the conventional channel conditions of the RF and FSO links and different values of $N$, there is no closed-form expression for (\ref{eq:eqWm1}). Thus, we use central limit theorem (CLT) to approximate $\mathcal{Y}_{(j,N)}$ by the Gaussian random variable $\mathcal{Z}_j\sim\mathcal{N}(\mu_j,\frac{1}{N}\sigma_j^2)$ where $\mu_j$ and $\sigma_j^2$ are the mean and variance derived based on the FSO channel condition.
%Reviewing the literature and depending on the channel condition, the FSO link is commonly considered to follow exponential, log-normal or Gamma-Gamma distributions, e.g., \cite{5342330,6168189,FSObook}.
For the Gamma-Gamma distribution of the FSO link \cite{5342330,6168189}, the channel gain follows the probability distribution function (PDF)
\begin{align}\label{eq:eqpdfgammagamma}
f_{G_\text{FSO}}(x)=\frac{2(ab)^{\frac{a+b}{2}}}{\Gamma(a)\Gamma(b)}x^{\frac{a+b}{2}-1}K_{a-b}\left(2\sqrt{abx}\right),
\end{align}
with $K_n(\cdot)$ denoting the modified Bessel function of the second kind of order $n$ and $\Gamma(x)=\int_0^\infty{t^{x-1}e^{-t}\text{d}t}$ being the Gamma function. Moreover, $a$ and $b$ are distribution shaping parameters.
 %which can be expressed as functions of Rytov variance \cite{6168189}.
%i.e., $f_{G_\text{FSO}}(x)=\lambda e^{-\lambda x}$ with $\lambda$ being the long-term channel coefficient, we have
In this way, denoting the expectation operator by $E\{\cdot\}$, the mean and variance of $\mathcal{Z}_j$ are found as
\begin{align}\label{eq:mueq}
&\mu_j=E\{\log(1+P_{\text{FSO},j}G_\text{FSO})\}=\frac{2(ab)^{\frac{a+b}{2}}}{\Gamma(a)\Gamma(b)}\int_0^\infty{x^{\frac{a+b}{2}-1}K_{a-b}\left(2\sqrt{abx}\right)\log(1+P_{\text{FSO},j}x)\text{d}x}
\end{align}
and $\sigma_j^2=\rho_j^2-\mu_j^2$ with
\vspace{-3mm}
\begin{align}\label{eq:sigmaeq}
&
\rho_j^2=E\{\log(1+P_{\text{FSO},j}G_\text{FSO})^2\}=
\frac{2(ab)^{\frac{a+b}{2}}}{\Gamma(a)\Gamma(b)}\int_0^\infty{x^{\frac{a+b}{2}-1}K_{a-b}\left(2\sqrt{abx}\right)(\log(1+P_{\text{FSO},j}x))^2\text{d}x},
\end{align}
which can be found numerically, because they are one-dimensional integrations.

Having $\mu_j$ and $\sigma_j^2$, the probabilities $\phi_j$ can be found for different RF channel models. Consider Rayleigh conditions for the RF link where $f_{G_\text{RF}}(x)=e^{-x}.$ Using (\ref{eq:eqWm1}) and $\mu_j$ and $\sigma_j^2$ in (\ref{eq:mueq})-(\ref{eq:sigmaeq}), the probabilities $\phi_j, \forall j,$ are given by
%\begin{align}\label{eq:cdfWmapprox}
%\phi_j&=\Pr(\log(1+P_{\text{RF},j}G_{\text{RF},j})+\mathcal{Y}_{(j,N)}\le R)\nonumber\\&=\int_0^{\frac{e^R-1}{P_{\text{RF},j}}}{f_{G_\text{RF}}(x)\Pr(\mathcal{Y}_{(j,N)}\le R-\log(1+P_{\text{RF},j}x))\text{d}x}\nonumber\\&\mathop  = \limits^{(a)}\int_0^{\frac{e^R-1}{P_{\text{RF},j}}}{e^{-x}Q\left(\frac{\sqrt{N}(\log(1+P_{\text{RF},j}x)+\mu_j-R)}{\sigma_j}\right)\text{d}x},\forall N,
%\end{align}
\vspace{-3mm}
\begin{align}\label{eq:cdfWmapprox}
\phi_j&=\int_0^{\frac{e^R-1}{P_{\text{RF},j}}}{f_{G_\text{RF}}(x)\Pr(\mathcal{Y}_{(j,N)}\le R-\log(1+P_{\text{RF},j}x))\text{d}x}\nonumber\\&\mathop  = \limits^{(a)}\int_0^{\frac{e^R-1}{P_{\text{RF},j}}}{e^{-x}Q\left(\frac{\sqrt{N}(\log(1+P_{\text{RF},j}x)+\mu_j-R)}{\sigma_j}\right)\text{d}x},\forall N,
\end{align}
where $(a)$ comes from the cumulative distribution function (CDF) of Gaussian distributions and CLT.
%Also, for the exponential, log-normal and the Gamma-Gamma distribution of the FSO link the mean and variance $(\mu,\sigma^2)$ are given by (\ref{eq:mueq})-(\ref{eq:sigmaeq}), (\ref{eq:eqmulognormal})-(\ref{eq:eqsigmalognormal}) and (\ref{eq:eqmugamma1})-(\ref{eq:eqsigmagamma1}), respectively. Therefore,
In this way, the final step to derive the throughput and the outage probability is to find (\ref{eq:cdfWmapprox}).
% while it does not have closed-form expression.
%he following lemmas propose \textcolor{blue}{several approximation/bounding} approaches for the CDF of the AMIs and, consequently, the throughput/outage \textcolor{blue}{probability.}
%\emph{\textbf{Lemma 1}}: The throughput and the outage probability of the HARQ-based RF-FSO setup are approximately given by
%\begin{align}\label{eq:eqapproxeta1}
%\eta=R\frac{1-\mathcal{F}(\frac{R}{M})}{1+\sum_{m=1}^{M-1}{\mathcal{F}(\frac{R}{m})}}
%\end{align}
%and
%\begin{align}\label{eq:eqapproxoutprob1}
%\Pr(\text{Outage})=\mathcal{F}\left(\frac{R}{M}\right),
%\end{align}
%respectively, with $\mathcal{F}(x)$ defined in (\ref{eq:cdfWmapproxlemma}).
%\begin{proof}
Therefore,
%for which
we implement the approximation $Q\left(\frac{\sqrt{N}(\log(1+P_{\text{RF},j}x)+\mu_j-R)}{\sigma_j}\right)\simeq V_{\alpha_j,\beta_j}(x)$ where
\vspace{-3mm}
\begin{align}\label{eq:linearizationnik2}
&V_{\alpha_j,\beta_j}(x)= \left\{\begin{matrix}
1 & x< \alpha_j-\frac{1}{2\beta_j},\\
\frac{1}{2}-\beta_j(x-\alpha_j) & x\in\Big[\alpha_j-\frac{1}{2\beta_j},\alpha_j+\frac{1}{2\beta_j}\Big],\\
0 & x> \alpha_j+\frac{1}{2\beta_j},
\end{matrix}\right.\vspace{-2mm}
\end{align}
with $\beta_j\doteq-\frac{\sqrt{N}P_{\text{RF},j}e^{\mu_j-R}}{\sigma_j\sqrt{2\pi}}$ and $\alpha_j\doteq\frac{e^{R-\mu_j}-1}{P_{\text{RF},j}},$
leading to
\vspace{-2mm}
\begin{align}\label{eq:cdfWmapproxlemma}
\phi_j&\simeq \int_0^{r_j}{e^{-x}V_{\alpha_j,\beta_j}(x)\text{d}x}= \int_0^{c_{1,j}}{e^{-x}\text{d}x}+\int_{c_{1,j}}^{c_{2,j}}{e^{-x}\left(\frac{1}{2}-\beta_j(x-\alpha_j)\right)\text{d}x}\nonumber\\&
=1-e^{-c_{1,j}}+e^{-{c_{2,j}}}\left(\beta_j {c_{2,j}}+\beta_j-\beta_j\alpha_j-\frac{1}{2}\right)
-e^{-{c_{1,j}}}\left(\beta_j {c_{1,j}}+\beta_j-\beta_j\alpha_j-\frac{1}{2}\right).
\end{align}
Here, $r_j\doteq\frac{e^R-1}{P_{\text{RF},j}},$ $c_{1,j}\doteq{\max\left(0,\alpha_j-\frac{1}{2\beta_j}\right)},$ $c_{2,j}\doteq{\min\left(\alpha_j+\frac{1}{2\beta_j},r_j\right)}$ and $V_{\alpha_j,\beta_j}(x)$ is obtained by applying Taylor expansion on the $Q$ function of (\ref{eq:cdfWmapprox}) at point $x=\alpha_j$.
Note that (\ref{eq:cdfWmapproxlemma}) is a tight approximation for moderate/large $N$'s for which CLT and (\ref{eq:linearizationnik2}) provide tight approximations. 

On the other hand, for Rician channel model of the RF link, which is of interest in line-of-sight conditions, the channel amplitude $\sqrt{G_\text{RF}}$ and gain ${G_\text{RF}}$, respectively, follow the PDFs
\begin{align}\label{eq:eqRicianpdf}
\tilde f_{\text{RF}}(x)=\frac{x}{\omega}e^{-\frac{(x^2+\nu^2)}{2\omega^2}}I_0\left(\frac{x\nu}{\omega^2}\right),
\end{align}
and $f_{G_\text{RF}}(x)=\frac{1}{2\sqrt{x}}\tilde f_{\text{RF}}(\sqrt{x})$, where $\nu$ and $\omega$ denote the fading parameters and $I_0$ is the zero-th order modified Bessel function of the first kind. In this way, $\phi_j$ is rephrased as
\begin{align}\vspace{-2mm}
&\phi_j=\int_0^{r_j}{\frac{\tilde f_{\text{RF}}(\sqrt{x})}{2\sqrt{x}}Q\left(\frac{\sqrt{N}(\log(1+P_{\text{RF},j}x)+\mu_j-R)}{\sigma_j}\right)\text{d}x}
\nonumber\\
&
=\int_0^{r_j^2}{{\tilde f_{\text{RF}}(u)}Q\left(\frac{\sqrt{N}(\log(1+P_{\text{RF},j}u^2)+\mu_j-R)}{\sigma_j}\right)\text{d}u}
\nonumber
\end{align}
\begin{align}\label{eq:eqbessel0}
&
\mathop  \simeq \limits^{(b)}\int_0^{r_j^2}{{\tilde f_{\text{RF}}(u)}V_{\tilde\alpha_j,\tilde\beta_j}(u)\text{d}u}
=\int_0^{\tilde c_1}{{\tilde f_{\text{RF}}(u)}\text{d}u}+\int_{\tilde c_1}^{\tilde c_2}{\tilde f_{\text{RF}}(u)(\frac{1}{2}-\tilde\beta_j(u-\tilde\alpha_j))\text{d}u}
\nonumber\\&
\mathop  \simeq \limits^{(c)}{\tilde F_{\text{RF}}(\tilde c_1)}+\left(\frac{1}{2}+\tilde\beta_j\tilde\alpha_j\right)\left({\tilde F_{\text{RF}}(\tilde c_2)}-{\tilde F_{\text{RF}}(\tilde c_1)}\right)\nonumber\\&-\tilde\beta_j\left(\tilde c_2\tilde F_{\text{RF}}(\tilde c_2)-\tilde c_1\tilde F_{\text{RF}}(\tilde c_1)-\left(\tilde c_2-\tilde c_1\right)\tilde F_{\text{RF}}\left(\frac{\tilde c_1+\tilde c_2}{2}\right)\right),
\end{align}
where $\tilde\alpha_j\doteq\sqrt{\frac{e^{R-\mu_j}-1}{P_{\text{RF},j}}},$ $\tilde\beta_j\doteq\sqrt{\frac{2NP_{\text{RF},j}e^{\mu_j-R}}{\pi}},$
$\tilde c_{1,j}\doteq{\max\left(0,\tilde \alpha_j-\frac{1}{2\tilde \beta_j}\right)},$ $\tilde c_{2,j}\doteq{\min\left(\tilde \alpha_j+\frac{1}{2\tilde \beta_j},r_j\right)}$. Also, $\tilde F_\text{RF}(x)=1-Q_{\mathcal{M}}\left(\frac{\nu}{\omega},\frac{x}{\omega}\right)$ is the CDF of the Rician variable (\ref{eq:eqRicianpdf}) with $Q_\mathcal{M}(\cdot,\cdot)$ being the Marcum $Q$ function, $(b)$ is based on the Taylor expansion of the $Q$ function and $(c)$ comes from the first order Riemann integral approximation $\int_{x_1}^{x_2}f(x)\text{d}x\simeq(x_2-x_1)f(\frac{x_1+x_2}{2})$.
\vspace{-0mm}
\subsection{Performance Analysis in the Cases with Comparable Coherence Times of the RF and FSO Links}
%Up to now, the results were presented for the cases with considerably shorter coherence time of the FSO link, compared to the RF link, motivated by the results of, e.g., \cite{5342330,598416,598420}.
%, such that the CLT provides accurate approximation for the sum of independent and identically distributed (IID) random variables.
%However, it is interesting t
Considering the cases with comparable coherence times of the RF and FSO links, i.e., with small values of $N$ in (\ref{eq:eqWm1}),
%Here, we mainly concentrate on the Gamma-Gamma distribution of the FSO link. The same results as in \cite{5357980} can be applied to derive the CDFs $F_{W_{m}}$ with, e.g., the exponential distribution of FSO link and small $N.$
%As seen in Figs. ?, the analytical results can be validated through simulations in the cases with exponential and log-normal distributions of the FSO link. However, to the best of authors' knowledge, there is no way to run the simulations for the cases with Gamma-Gamma distribution of the FSO link fading variables in (\ref{eq:eqWm1}). Therefore, we compare the analytical results of the Gamma-Gamma based condition with the upper and lower bounds given in the following.
we use Minkowski inequality \cite[Theorem 7.8.8]{minkowskibook}
\begin{align}\label{eq:eqminko}
\left(1+\left(\prod_{i=1}^n{x_i}\right)^\frac{1}{n}\right)^n\le \prod_{i=1}^{n}{\left(1+x_i\right)},
\end{align}
to write
\vspace{-2mm}
\begin{align}\label{eq:eqminkowseq1}
&\Pr\left(\mathcal{Y}_{(j,N)}\le x\right)= \Pr\left(\prod_{i=1}^N(1+P_{\text{FSO},j}G_{\text{FSO},1+(j-1)i})\le e^{Nx}\right)\nonumber\\&\le \Pr\left(1+P_{\text{FSO},j}\left(\prod_{i=1}^N{G_{\text{FSO},1+(j-1)i}}\right)^{\frac{1}{N}}\le e^{x}\right)=F_\mathcal{Q}\left(\left(\frac{e^x-1}{P_{\text{FSO},j}}\right)^{N}\right),
\end{align}
where using the results of \cite[Lemma 3]{6168189} and for the Gamma-Gamma distribution of the variables $G_{\text{FSO},1+(j-1)i}$, $\mathcal{Q}=\prod_{j=1}^m\prod_{i=1}^N{G_{\text{FSO},1+(j-1)i}}$ follows the CDF
\begin{align}\label{eq:eqresminkows}
F_\mathcal{Q}(x)=\frac{1}{\Gamma^{N}(a)\Gamma^{N}(b)}\mathcal{G}_{1,2N+1}^{2N,1}\Bigg(({ab})^{N}x\Bigg|_{\underbrace{a,a,\ldots,a}_{N \text{ times}},\,\underbrace{b,b,\ldots,b}_{N \text{ times}},0}^{\,\,\,\,\,\,\,\,1}\Bigg),
\end{align}
with $\mathcal{G}(.)$ denoting the Meijer G-function.

In this way, from (\ref{eq:eqWm1}) and (\ref{eq:eqresminkows}), the probabilities $\phi_j,$ are tightly bounded by
\begin{align}\label{eq:equpperboundG}
&\phi_j\le  \frac{1}{\Gamma^{N}(a)\Gamma^{N}(b)}\int_0^{r_j}{e^{-x}\mathcal{G}_{1,2N+1}^{2N,1}\Bigg(}{\bigg(\frac{ab(e^{R-\log(1+P_{\text{RF},j}x)}-1)}{ P_{\text{FSO},j}}\bigg)^{N}\Bigg|_{\underbrace{a,a,\ldots,a}_{N \text{ times}},\,\underbrace{b,b,\ldots,b}_{N \text{ times}},0}^{\,\,\,\,\,\,\,\,1}\Bigg)\text{d}x}
\end{align}
which
%, because it is a one-dimensional integration,
can be calculated numerically (see Section IV for the tightness of approximations). Also, we can replace $f_{G_\text{RF}}(x)=\frac{1}{2\sqrt{x}}\tilde f_{\text{RF}}(\sqrt{x}),$ ($\tilde f_{\text{RF}}({x})$ given in (\ref{eq:eqRicianpdf})) into (\ref{eq:equpperboundG}) to derive the results for the Rician RF links.
Finally, note that the results of (\ref{eq:equpperboundG}) is mathematically applicable for every value of $N$. However, for, say $N\ge 6,$ the implementation of the Meijer G-function in MATLAB is very time-consuming.
%and the tightness of the approximations (\ref{eq:equpperboundG}), (\ref{eq:eqlowermolla}) and (\ref{eq:eqxxxx}) decreases for large values of $N$.
 As a result, (\ref{eq:equpperboundG}) is useful for the performance analysis in the cases with small $N$'s, while the CLT-based approach provides accurate performance evaluation as $N$ increases.
\subsection{On the Effect of Adaptive Power Allocation}
As shown in, e.g. \cite{MIMOARQkhodemun}, adaptive power allocation between the ARQ retransmissions  leads to marginal throughput increment, while the outage probability indeed benefits substantially from optimal power allocation. If the data retransmission stops at the end of the $m$-th round, the total consumed energy is $\xi_{(m)}=L\sum_{j=1}^mP_{j}, P_j=P_{\text{RF},j}+P_{\text{FSO},j}$ where $L$ is the length of the codewords. Thus, following the same procedure as in \cite{6918481}, the normalized expected consumed energy (normalized by the length of the codewords) is found as
\vspace{-2mm}
\begin{align}\label{eq:eqnormalizedenergy}
\Xi=\sum_{m=1}^M{P_m\Phi_{m-1}}.
\end{align}
To optimize the power allocation, in terms of outage probability, we need to use (\ref{eq:probphim}), (\ref{eq:eqnormalizedenergy}), $\Pr(\text{Outage})=\Phi_M$ and a Lagrange multiplier criteria to optimize the power terms by setting the derivatives of the criteria equal to zero. However, due to the complex expressions of the probabilities $\phi_j,\forall j,$ it is difficult to follow the derivative-based approach. Instead, we propose a suboptimal scheme where the expected consumed energy in each retransmission is set to be the same, i.e., $P_m\Phi_{m-1}=P_n\Phi_{n-1}, \forall m,n$. The intuition behind the considered power allocation is to weight the energy in each round by its consumption probability, such that more energy is assigned to the last retransmissions, which are rarely used.

With no loss of generality, let us assume $P_{\text{RF},j}=P_{\text{FSO},j},\forall j,$ while the same discussions hold for other relations between $P_{\text{RF},j}, P_{\text{FSO},j}.$ Using $P_m\Phi_{m-1}=P_n\Phi_{n-1}, \forall m,n$, (\ref{eq:eqnormalizedenergy}) and an energy-per-codeword budget $\Xi=\bar P$ we have $P_1=\frac{\bar P}{M}$ and the other power terms are
%all power terms, and correspondingly, the outage probability and the expected energy (\ref{eq:eqnormalizedenergy}) can be
expressed only as a function of $P_1$ via
\begin{align}\label{eq:eqheurispower}
\,\,\,\,\,&P_n={c}\bigg(\prod_{j_0=1}^{n-2}\bigg(\mathcal{F}\bigg({c}{\bigg(\prod_{j_1=1}^{j_0-1}\mathcal{F}\bigg({c}{\bigg(\prod_{j_2=1}^{j_1-1}\mathcal{F}\bigg(\frac{c}{\prod_{j_3=1}^{j_2-1}\ldots}\bigg)\bigg)}\bigg)\bigg)^{-1}}\bigg)\bigg)^{-1}\bigg),
 n\ge 2, c=\frac{P_1}{\mathcal{F}(P_1)},
\end{align}
%\begin{align}\label{eq:eqheurispower}
%P_n=\frac{c}{\prod_{j_0=1}^{n-2}F\left(\frac{c}{\prod_{j_1=1}^{j_0-1}F\left(\frac{c}{\prod_{j_2=1}^{j_1-1}F\left(\frac{c}{\prod_{j_3=1}^{j_2-1}\ldots}\right)}\right)}\right)}.
%\end{align}
where $\mathcal{F}(P)$ denotes evaluation of $\phi_j$ for power $P=P_\text{RF}+P_\text{FSO}$.
%Finally, note that the intuition behind the considered power allocation is to weight the energy in each round by its consumption probability, such that more energy is assigned to the last retransmissions, which are rarely used.
\vspace{-3mm}
\section{Numerical Results and Conclusions}
In all figures, we set $\alpha=4.3939, \beta=2.5636$ which correspond to Rytov
variance 1 of the FSO link \cite{6168189}. Also, the parameters of Rician RF PDF in (\ref{eq:eqRicianpdf}) are set to $\omega=0.7036, \nu=0.0995,$ leading to unit mean and variance of the channel gain distribution $f_{G_\text{RF}}(x)$. In Figs. 1a-c, we consider a peak power constraint $P_{\text{RF},j}=P_{\text{FSO},j}=P,\forall j,$ for the RF and FSO links. In Fig. 1d, however, we present the results for the cases with a normalized expected energy constraint $\Xi=\bar P$.

\begin{figure}
\centering
  \includegraphics[width=0.55\columnwidth]{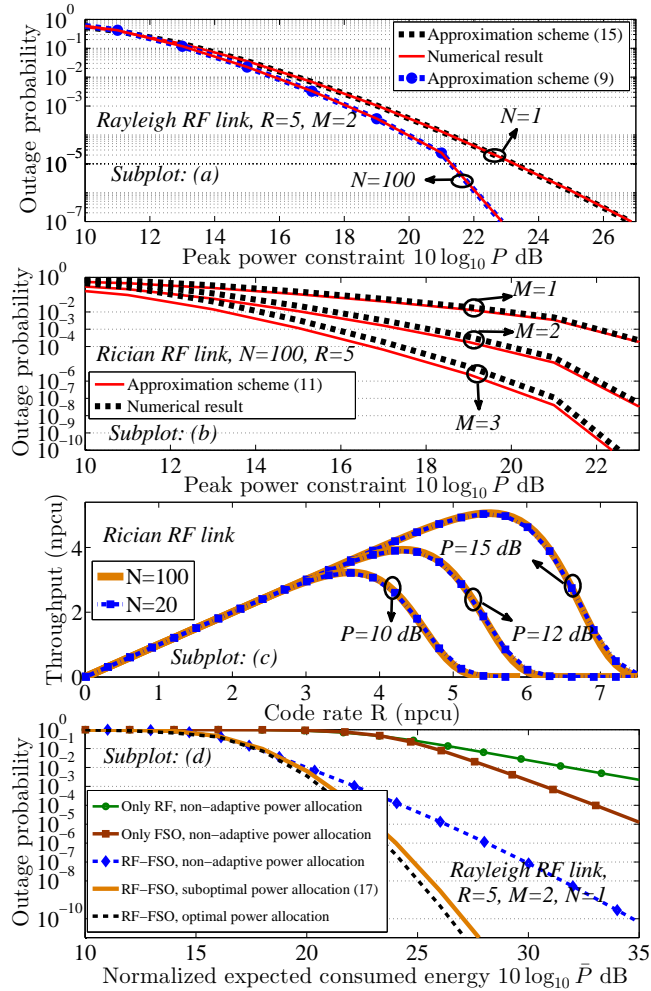}\\\vspace{-5mm}
\caption{(a)-(b): Outage probability vs peak power constraint $P_{\text{RF},j}=P_{\text{FSO},j}=P,\forall j,$ in Rayleigh and Rician RF links. (c): Throughput vs the code rate $R$, Rician RF link. (d): Outage probability vs the normalized expected consumed energy constraint $\Xi= \bar P,$ Rayleigh RF link. }\label{figure111}\vspace{-6mm}
\end{figure}
Figures 1a-b study the outage probability for Rayleigh and Rician channel models of the RF link, respectively, and investigate the tightness of the proposed approximation schemes. Then, Fig. 1c evaluates the throughput in the cases with Rician RF link and different numbers of channel realizations in the FSO link $N$. Note that, with non-adaptive power allocation, we have $\phi_i=\phi_j,\forall i,j>0$ (see (\ref{eq:cdfWmapprox})), leading to $\eta=R(1-\phi_1)$ in (\ref{eq:eqeta1}). Thus, the results of Fig. 1c are independent of the number of retransmissions $M$. Finally, Fig. 1d studies the effect of the proposed suboptimal and optimal (optimized by exhaustive search) power allocation between the retransmissions, and compares the system performance with the cases using only the RF or the FSO link. In the meantime, we have checked the results for other parameter settings which, due to space limits, are not reported in the figures. According to the figures, the following conclusions can be drawn:

%\begin{itemize}
1) The approximation approaches of Sections III. A and B are very tight for a broad range of parameter settings, and depending on the relative coherence times of the links there are different methods for the performance analysis of RF-FSO systems. (Figs. 1a-b).

2) The use of ARQ reduces the outage probability remarkably, compared to the cases without ARQ, i.e., $M=1$ (Fig. 1b). However, the throughput is not affected, if the power terms are not adapted in the retransmissions (Fig. 1c).

3) For small values of $R$, the throughput increases with the rate (almost) linearly, because with high probability the data is correctly decoded in the first round. On the other hand, the outage probability increases and the throughput goes to zero for large values of $R$.

4) The RF-FSO link leads to substantially less outage probability, compared
to the cases with only the RF or the FSO link. Intuitively, this is because with the joint RF-FSO setup the diversity increases and the RF (resp. the FSO) link compensates the effect of the FSO (resp. RF) link, if it experiences poor channel conditions (Fig. 1d).

5) Finally, adaptive power between the retransmissions reduces the outage probability significantly, and the proposed suboptimal power allocation scheme mimics the optimal power allocation very tightly (Fig. 1d).
\vspace{-3mm}
\bibliographystyle{IEEEtran} %lic.bst is the style file
\bibliography{masterFSOletter}
\vfill
\end{document}